\documentclass[prl,twocolumn,superscriptaddress,groupedaddress]{revtex4}  
\usepackage{graphicx}  
\usepackage{dcolumn}   
\usepackage{bm}        
\usepackage{amssymb}   
\usepackage{fancybox}
\usepackage{cases}
\usepackage{subfigure}
\begin{document}
\title{Comments on ``Unusual Thermoelectric Behavior Indicating a Hopping to Bandlike Transport
Transition in Pentacene''} \maketitle

{\vspace*{-9mm}}W. Chr. Germs, K. Guo, R. A. J. Janssen, and M.
Kemerink {\cite{germs}} recently measured the temperature and
concentration dependent seebeck coefficient in organic thin film
transistor and found the seebeck coefficient increases with carrier
concentration (corresponding to the gate voltage) in the low
temperature regime. They further concluded that this unusual
behavior was due to a transition from hopping transport in static
localized states to bandlike transport, occurring at low
temperature. This is obviously in contrast to the previous
theoretical prediction because it is widely accepted that hopping
transport is more pronounced at low temperature. We will discuss the
reason for this unusual behavior here and suggest that the density
of states function plays an
important role in concentration dependent seebeck coefficient.\\
Note that, based on the percolation theory and hopping transport,
the seebeck coefficient $S$ could be calculated as \cite{tri,
visenberg} \vspace*{-3mm}\begin{equation}
S=\frac{k_B}{e}\frac{\int\frac{E_i+E_j-2 E_F}{2k_BT}
g\left(E_i\right)g\left(E_j\right)dE_idE_jd{\bf{R}}\theta\left({s_c-s_{ij}}\right)}{\int
g\left(E_i\right)g\left(E_j\right)dE_idE_jd{\bf{R}}\theta\left({s_c-s_{ij}}\right)}.
\end{equation}
Where $g\left(E\right)$ is the density of states (DOS) at energy
$E$, $s_{ij}=2\alpha R_{ij}+\frac{\mid E_i-E_F\mid+\mid
E_j-E_F\mid+\mid E_i-E_j\mid}{2k_BT}$ according to the hopping
theory, and $s_c$ is the critical percolation parameter
\cite{visenberg}. In organic semiconductor, DOS is usually described
as single Gaussian or exponential function, the seebeck coefficient
in this situation is found decreasing with carrier concentration
based on equation (1). However, DOS in organic semiconductor is very
complicated and single Gaussian or exponential function can
approximated only part of the density distribution and the different
DOS function will affect the seebeck coefficient. For example, if we
choose the DOS as function as
$g\left(E\right)\propto\left(1+\left(E_0+\mid
E\mid\right)^{-p}\right)$($p>0$, which shows good fitting with
measured deep energy DOS for organic semiconductor \cite{exp}(inset
of Fig.1 (a)) including pentacene \cite{exp2}), in this situation,
both conductivity and seebeck coefficient increase with fermi energy
as shown in Fig. 1 (a)($s_c$ decreasing means conductivity
increasing and fermi energy is
corresponding to carrier concentration or gate voltage).\\
Next, we want to interpret this behavior. In fact, the seebeck
coefficient could be simplified as  $\mid S\mid\propto
\frac{dLog\left(\mu\left(E\right)g\left(E\right)\right)}{dE}$
\cite{dos}, it is very clear that the energy differential of the DOS
affects the S-value. At high temperature,  the transport path of
carrier lies in the DOS for shallow energy ($E>E^*$), $dLog(DOS)/dE$
monotonically decreases with energy and the S-value  decreases
accordingly with concentration (carriers will occupy the higher
energy with concentration increasing); however, at low temperature,
 the carrier transport path is in the deep energy states as shown in Fig.1 (b), if
$dLog(DOS)/dE$ increases with energy, the S-value is no surprised to
possibly increase with concentration. Certainly, the relation
between $\mu$ and $E$ should be also taken into account, as we have
done in Fig. 1(a).Moreover,the same conclusion is reached if we
substitute the DOS in Fig.1 (a) and transport energy model
\cite{li}into equation (1) in \cite{germs}.
\begin{figure}
\centering \subfigure[]{
\label{fig:subfig:a} 
\includegraphics[width=2.2in]{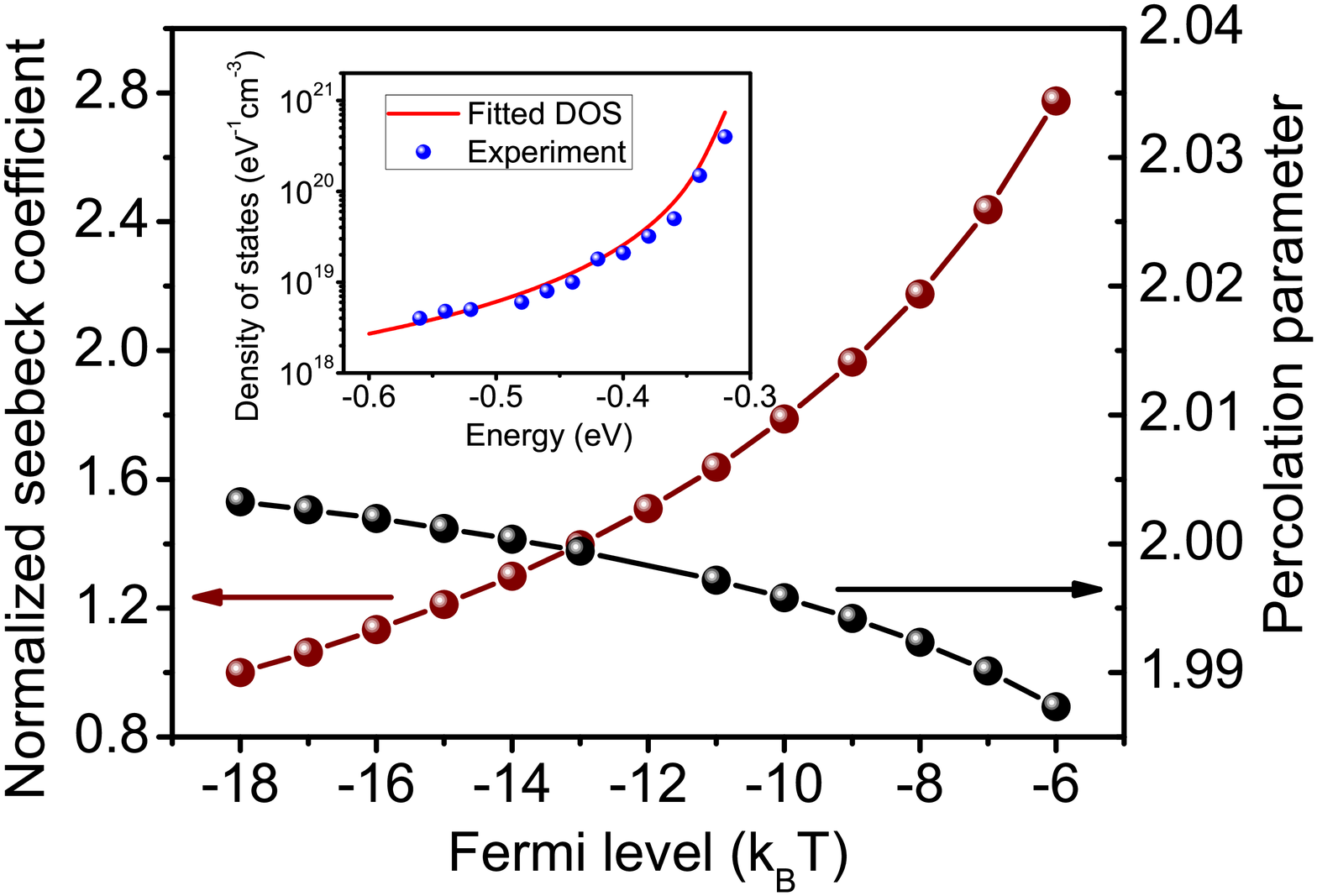}}
\hspace{1in} \subfigure[]{
\label{fig:subfig:b} 
\hspace*{0.8cm}\includegraphics  [width=2.1in]{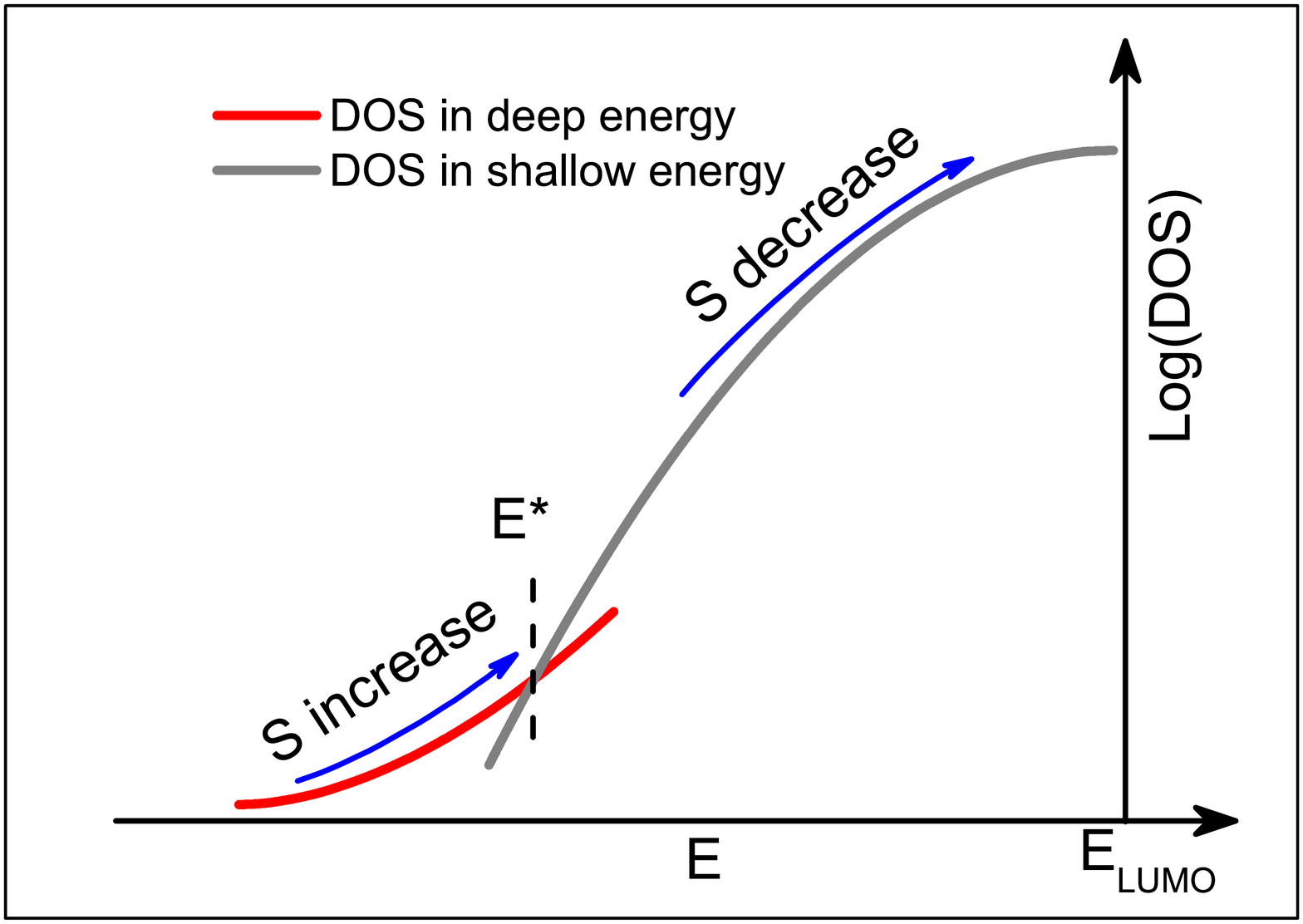}}
\vspace*{-2mm}\caption{(a) The calculated fermi level dependent
percolation parameter $s_c$ and normalized seebeck coefficient;
Insert: Comparison between fitted (line) and measured DOS(symbols).
(b) Schematic DOS-$E$ model for this work.}
\label{fig:subfig} 
\end{figure}

Hence we conclude that, the seebeck coefficient increases with
concentration in \cite{germs} might be attributed to the DOS shape
in the low temperature, and this experiment could not prove hopping transport to be invalid in pentacene transistors.\\
Ling Li,  Nianduan Lu, and Ming Liu.\\
 Institute of
microelectronics, Chinese Academy of Sciences, Beijing, China,
100029. 


\vspace*{-8mm}
\begin{thebibliography}{99}
\vspace*{0mm}
\bibitem{germs}
W. Chr. Germs et al., \emph{Phy. Rev. Lett} {\bf 109}, 016601
(2012).
\bibitem{tri}
G. P. Triberis et al., \emph{J. Non-Cry. Solids}. {\bf 79}, 29
(1986); M. Grunewald, and P. Thomas, \emph{Phys. Stat. Sol}. (b)
{\bf 93}, K17 (1979).
\bibitem{visenberg}
M. C. J. M .Vissenberg et al., \emph{Phy. Rev. B} {\bf 57}, 12964
(1998);
\bibitem{exp}
 W.S.C.Roelofs et al., \emph{Phy. Rev. B}, {\bf 85},
085202 (2012);
\bibitem{exp2}
S.Yogev et al., \emph{Phy. Rev. B}, {\bf 84}, 165124(2011);.
\bibitem{dos}
 H. Fritzsche, \emph{Sol. Stat. Comm.}, {\bf{9}}, 1813 (1971).
\bibitem{li}
 L. Li et al.,\emph{Appl. Phy. Lett.}, {\bf 92},013307 (2008).
\end{thebibliography}
\end{document}